# Emergence of Collective Memories

**Sungmin Lee** IceLab, Department of Physics, Umeå University, 90187 Umeå Sweden
**Veronica C Ramenzoni** Department of Psychology, University of Virginia, Charlottesville, VA 29003, USA
**Petter Holme** IceLab, Department of Physics, Umeå University, 90187 Umeå Sweden and Department of Energy Science, Sungkyunkwan University, Suwon 440–746, Korea

**Background:** We understand the dynamics of the world around us as by associating pairs of events, where one event has some influence on the other. These pairs of events can be aggregated into a web of memories representing our understanding of an episode of history. The events and the associations between them need not be directly experienced—they can also be acquired by communication. In this paper we take a network approach to study the dynamics of memories of history.

**Methodology / Principal Findings:** First we investigate the network structure of a data set consisting of reported events by several individuals and how associations connect them. We focus our measurement on degree distributions, degree correlations, cycles (which represent inconsistencies as they would break the time ordering) and community structure. We proceed to model effects of communication using an agent-based model. We investigate the conditions for the memory webs of different individuals to converge to collective memories, how groups where the individuals have similar memories (but different from other groups) can form.

**Conclusions / Significance:** Our work outlines how the cognitive representation of memories and social structure can co-evolve as a contagious process. We generate some testable hypotheses including that the number of groups is limited as a function of the total population size.

## Introduction

When we think and talk about changes in the world around us, we weave together discrete episodes to create a narrative. Events along the timeline of this narrative are connected by associations [1,2]. For instance, if we recall that "since the gas price went up, I decided to buy a fuel-efficient car", we can represent "gas price going up" and "I buying a fuel-efficient car" as two nodes of a directed graph with an associative arc connecting the former to the latter. Such arcs need not be a first-hand experience; we might believe that "increased political tension in the middle-East drove the gas price up" drawing an arc from "increased political tension in the middle-East" to "gas price going up". Our recollection and understanding of the past can thus be represented as a web of events connected by associations like the above example. Such autobiographical narratives take place within a social context [3,4]. They are fluid and dynamic, bearing the hallmark of the social context within which they emerge. Associations between events are not only dependant on the personal experiences of the individual, but also on social processes of construction and re-construction that ultimately give place to what we experience as a collective history. Much of the social processes of collective history occur in conversational and communicative contexts, which convey personal and social meaning to events. Communication reinforces the memories of interacting individuals, and it is through this process that associative arcs can spread in a population so that memory webs come to share common elements across people. From this process, associative arcs can spread in a population so that groups of people share part of their webs of memories. We call such subnetworks in common to many people *collective memories*. In this paper, we will investigate a model of how the web of memories of a population evolves, including the mechanisms mentioned above. We use the model to investigate the stability of collective memories, the minimal requirements for cycles (sequences of associations that must violate the time-ordering of the events) and the possibility that communication can lead to the formation of groups sharing collective memories. There are other conceivable mechanisms than communication for the evolution of a person's web of memories: firsthand experiences, mass-medial information and logical deduction (to fill out gaps in one's web of memories). In our model of the dynamics of collective memories these other mechanisms are grouped together and, as opposed to communication, treated as external input to the model.

Communications as a process for the understanding of history—the formation of causal networks, both on an individual and aggregate level—has been studied in the qualitative tradition of social and behavioral sciences. One has investigated the processes behind collective memories—how groups of people maintain a common narrative of a period of history. This type of research are to a large extent case studies about, ethnic groups' memories of traumatic events, like the Jews' collective memories of the holocaust [5,6], or comparative studies like the Palestinians and Israelis different histories of the state of Israel [7].

Recently there has been a considerable interest in models of the spreading of information and opinions between people. Such studies have for example investigated the minimal requirements for fads to spread [8], for groups of people to make correct collective decisions or predictions [9], and the conditions for a diversity of opinions (as opposed to a widespread consensus) to be the result of communication [10]. In the present work we follow this tradition and create a model of collective memory emerging from communication. This model will take a web of memories as input. In this study we take this input network from an empirical dataset. The paper starts discussing the structure of this empirical data, then proceeds to the construction of the model and finally discusses the results of the simulations.

## Results

### Network structure of aggregated memories

We start by characterizing the network structure of and empirical dataset of aggregated memory webs. We will later use this dataset as a prototype memory web in our simulation study. The dataset was collected from Ref. [11] by Bearman, Faris and Moody [12] who used it to propose a network aided method of assembling a case from heterogeneous interview data (rather than studying social memory processes). The source contains interviews of 14 people living in a village in northern China telling their life-stories. Many of these stories concern the period a few decades before the interviews, when armed conflicts between Chinese and Japanese troops and also by Nationalist and Communist Chinese affected the village. From these accounts Bearman et al. identified "narrative clauses" where the interviewee associates one event with another (for details, see Ref. [12]) and connected these into an aggregated memory web of 1995 vertices (events), about a fourth of these are common to two or more individuals, and 2121 association arcs.

The first quantities we studied were the in- and outdegrees. The outdegree of an event $v$, $k_{out}(v)$, is the number of other events $v$ is thought to affect, and thus a measure of how influential $v$ is. The indegree $k_{in}(v)$, is the number of other events thought to affect $v$, and thus a measure of how complex the causes of an event $v$ were. 281 of the vertices have zero indegree and 576 have zero outdegree. The boundary of the memory will appear as zero-degree vertices, but in interview-based data like ours there can also be other reasons for an event having no in- or out-neighbors. A zero-indegree vertex may represent an event perceived too general to need an explanation; a zero-outdegree vertex may represent an event too new to have influenced events important enough to mention. Most of the events are thus intermediate in the narrative chains of associations, both perceived as caused by some other event and causing other events.

Over the last decade there has been a tremendous interest in power-law degree distributions [13], where the frequency of a vertex with degree $k$, $p(k)$, is proportional to $k^{-\gamma}$, for $k > k_{min}$. This interest has several distinct reasons. First, power-law degree distributions are observed in a great diversity of systems and still are not explained by simple random processes. Another reason and partly because they have rather special statistical properties—essentially, power-law degree distributions are much more heterogeneous than normally distributed quantities like body height. (E.g., if the body height of Americans was power-law distributed with minimum cut-off of 4'5 and average 5'7, then there would be about 40,000 Americans taller than the Empire State Building.) A common way of detecting power-law distribution is to plot the cumulative mass function $P(k)$, the frequency of degrees larger than a particular value, on double logarithmic axes. The reason one plots the cumulative distribution rather than $p(k)$ is that if $p(k)$ is a power-law, then so is $P(k)$ (but with the exponent $-\gamma + 1$ rather than $-\gamma$) but the fluctuations in the tail is smaller; the reason for the double logarithmic axes is that a power-law appears as a straight line in such plots. In Fig. 1A we plot $P(k)$ for the in- and outdegree, and indeed they are rather straight lines (especially the $k_{in}$ distribution). Assuming the distributions are power-laws, we obtain the values of the exponents $\gamma_{in}$ = 3.17(6) and $\gamma_{out}$ = 2.7(2). However, more strict statistical test gives low probabilities that the observed data really come from power-law distributions (p-values less than 0.01 for both the in- and outdegree distributions respectively). For the indegree distribution, that fits a straight line very well, a power-law would give a few vertices with even larger degree (the expected largest degree for a degree distribution $k^{-3.17}$ and 1995 vertices is ~33). To sum, the degree distributions are skewed and broad—there are many minor events and a few very influential—but not power-laws.

To get a more detailed picture of the aggregate memory web we measured the clustering coefficient and assortativity (see Methods). The clustering coefficient quantifies the number of triangles where one event A is both directly associated with another B, and indirectly, via an event C, relative to the number of connected triples of events. A triangle can be a sign of individuals changing the resolution of their narrative—in a brief account A can be directly associated with C, but going into more detail B is inserted as a further explanation of the A–B association. The clustering coefficient for the whole network is 0.0336—much closer to the minimum, 0, than the maximum, 1. It is however much larger than what would be expected from a random reference model with the same set of degrees as the real memory web (see Methods), 0.0017 (s.e. 0.0001). Our conclusion is that the narratives at various resolution (as mentioned above) is present in the data, but this tendency is not strong enough to make the network markedly triangle rich. A yet more detailed picture of the triangle structure is given in Fig. 1B where we plot the local clustering coefficient (a quantity of individual vertices, see Methods). This is interesting because some studies [14] claim that if the clustering decays with the degree, then the network probably has a hierarchical structure (where the vertices of highest degree controls a level below, which in turn controls a level below, and so on.). In our case this is not true, the local clustering coefficient is almost independent of degree (only weakly decaying). A non-hierarchical organization would then imply that the most influential (in some sense) events are associated, albeit not primarily, with relatively influential events as well as less significant ones. If we replace "influential" by outdegree this statement can be measured directly by the *assortativity*. This measure quantifies the correlation between the degrees at the two sides of an arc. A negative value means that there is a tendency for low-outdegree vertices (events that are associated with few more recent events) to be connected to high-indegree vertices (events that are associated with many earlier events). In our data the assortativity is –0.00226, very close to 0 (that indicates neutrality). The condition that only one arc can go between two vertices in combination with a broad degree distribution can cause the assortativity to become negative in reference models as ours [15,16], but the difference is not so large as the average value over the null-model is –0.034(1). To get a more detailed picture we plot the average indegree (outdegree) of a neighbor, $k_{nn}$, as a function of the indegree (outdegree) and find this quantity to be even more degree independent than the clustering coefficient. There is not a strong correlation between the earlier and later event of an association arc.

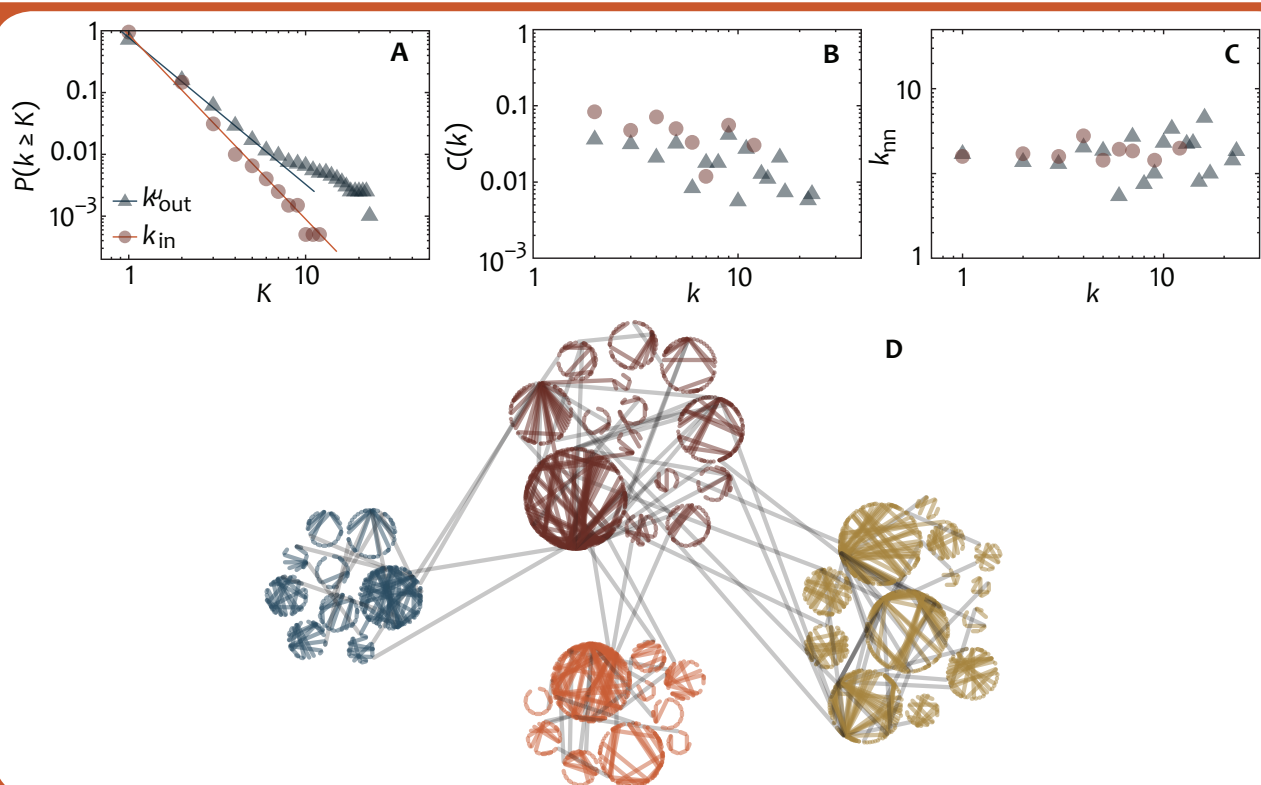

**Fig. 1.** Structural properties of an empirical aggregated memory web. (A) displays cumulative in- and outdegree distributions. (B) shows the clustering coefficient as a function of the in and outdegrees. In panel (C) we investigate the degree–degree correlation by plotting the average neighbor degree $k_{nn}$ as a function of the in- and outdegree of an agent. (D) is a plot of the largest connected component of the aggregated memory web, highlighting the modular structure. A weighted-network clustering scheme identifies 48 smaller groups. If these are treated as vertices and the same clustering scheme is applied to that network of groups, we discover four supergroups indicated by different colors.

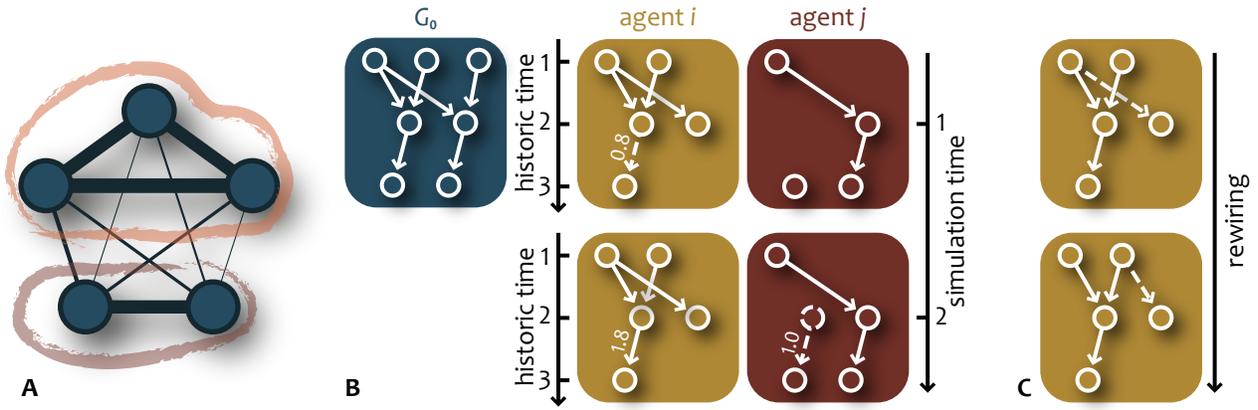

**Fig. 2.** Illustrations of the model. (A) shows an illustration of the social network of agents. The thickness of the lines are proportional to the strength of the social connection—the familiarity *F*. The social networks can have communities (encircled in the figure)—groups of agents that are more strongly connected (in terms of familiarity) within than between each other. (B) illustrates the communication between two agents *i* and *j*. The initial memory webs are derived from the same seed graph $G_0$. At simulation time step 1 agent *i* and *j* are selected to communicate (this selection is random with a probability proportional to the strength of their social ties). Then an association tie between two events is selected as a topic of communication between *i* and *j*. As a result of the communication, one event vertex and one arc with weight one is added to agent *j*'s memory web, whereas the corresponding arc of *i*'s memory web increases in weight by one unit. (C) illustrates the randomness introduced to the memory webs by rewiring of arcs.

As a final quantity to characterize the empirical network, we measure the maximal modularity *q*—a measure that reflects how well a graph can be divided into groups that are densely connected within and strongly connected between each other (see the Methods section). We measure this value to 0.918 to be compared with 0.847(1) for the reference model. So even if much of the high *q*-value can be explained by the degree sequence, there is a positive tendency for the events to occur in groups. The events are thus clustered into groups that can be thought of as super events. In Fig. 1D we plot the empirical memory web with the detected groups as circles. These groups are then clustered into four super clusters (clusters of clusters).

If we assume that the events are not overlapping in time, a cycle in the data (a closed chain of associative arcs) would be a contradiction (since at least one arc would correspond to one event influencing another that has already happened). In the real data there are indeed four such cycles between six and nine arcs in length. This means that the aggregate story either has contradictions (an interpretation implicit in Ref. [12]), or that there are some event A that is so long in duration that it can affect an event B that subsequently (directly or indirectly) affects the latter stages of A. Disambiguating this issue is beyond the scope of this paper. In our model of collective memory dynamics, we will assume that events do not non-overlap (so that a cycle is an inconsistency) and we will discuss this model against the alternative scenario, which allows for events to overlap.

**Communication model**

Now we turn to our model of collective memories. We will give an overview of the assumptions, structure and basic parameters in these sections; more technical details are provided in the Methods section.

Our model has three basic assumptions. First, that people, while communicating, influence each other's associations; so that the higher the frequency with which an individual hears another person make an association between events, the stronger that association becomes in the individual's memory web. Second, the stronger an association is, the more likely an individual is to talk about it. Third, people are more likely to communicate with people that they perceive as similar to them (in terms of age, interests, location, etc. [17]). Then we consider a population of $N$ individuals that communicate about history. Every agent $i$ has its own memory web $G_i$ of $i$ vertices and weighted arcs where the weight $w_i(u,v)$ represents the strength of user $i$'s association between the events $u$ and $v$. We also have the constraint that an individual agent's network does not have any cycles (which would then constitute inconsistencies).

What demarcates an event is a result of the physical events (the bullet leaving the gun of an assassin), social communication processes ("have you heard that Archduke Franz Ferdinand was killed?") and also cognitive processes of the individuals. Moreover, in narrating a historical episode, what constitutes an event is a consequence of the level of detail used (a coarse level "the murder of Archduke Franz Ferdinand sparked World War I" or a detailed level including a description of the role of the Triple Entente Powers in Europe's politics, etc.). Here, however, we assume, for the sake of simplicity, that the initial memory web is drawn from a common web of events $G_0$ (which are taken from the data mentioned above with the few cycles removed to increase data consistency). As a result, we assume that agents communicate about the events at the same level of generalization and that agents agree on how to delimit an event. These elements can be incorporated in a rather straightforward fashion and their potential impact on the outcome of this work is minimal.

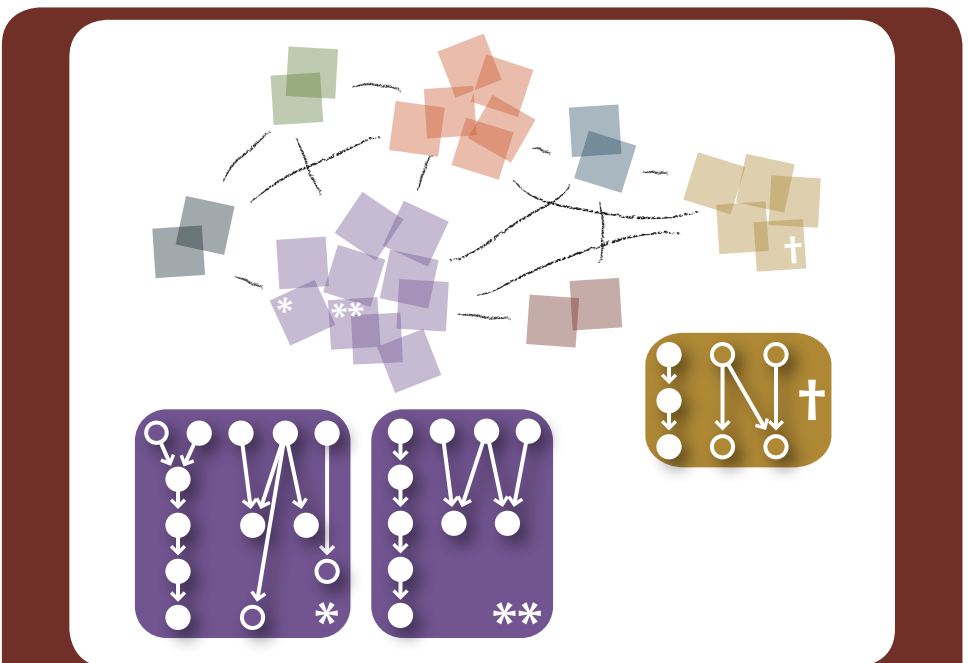

**Fig. 3.** Example configuration of the memory webs of the agents after a simulation with $N = 30$, $r = 30$, $\rho = 1$, and $\varphi = 0.3$. The group structure reflects the similarities of memory webs as described in the text. This is exemplified by the memory webs of three agents—two in the same cluster and one in another cluster.

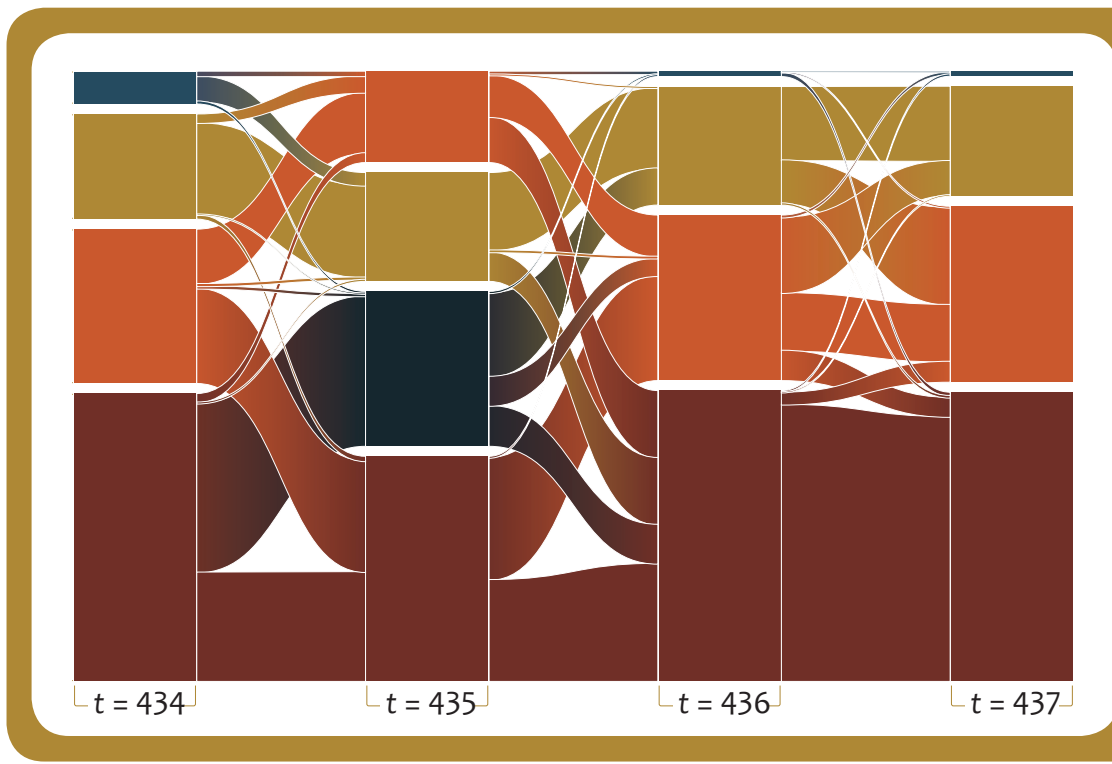

**Fig. 4.** Example of the merging and splitting of memory groups in a run with parameter values with $N = 1000$, $r = 5000$, $\rho = 0$ and $\varphi = 0.3$.

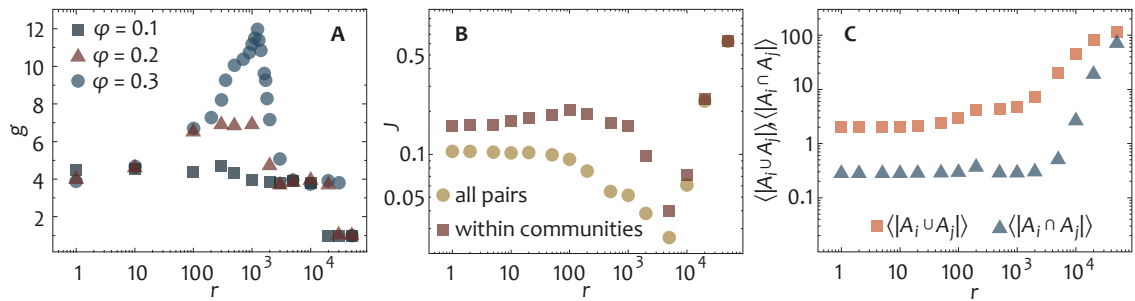

**Fig. 5.** Group-size and memory-similarity statistics for systems with 1000 agents and ρ = 0 at the final time step of the simulation t = 440, we use averages over 50 runs of the algorithm. (A) shows the number of detected groups g against communication rate r for φ = 2.1, 0.2, and 0.3. (B) displays the Jaccard index measure of similarity averaged over all pairs of vertices and pairs of vertices in the same community of the social network. In (C) we graph the average value of J's components—$|A_i \cap A_j|$ and $|A_i \cup A_j|$ against r for φ = 0.3—for all pairs. We omit errorbars as they are smaller than the symbol size.

The simulation proceeds by random pairs of agents communicating about the history and learning from each other. At a time step, we pick an agent with uniform randomness and choose, r times, both someone to talk with and something to talk about. Each pair of agents are associated with a familiarity score F that represents the strength of their social relationship. Given an agent i, a communication partner j is chosen randomly with a probability proportional to $F_{ij}$. As a result, agents that are more closely acquainted will influence each other more frequently than agents that are less closely acquainted. The initial value of F is 1. During a time step, if i and j communicate, $F_{ij}$ increases by 1. The dynamics of the memory webs follow a similar reinforcement dynamics as the familiarities. When choosing i and j (as described above) a random arc (u,v) of i's memory web with a probability proportional to the weight $w_i(u,v)$ was also chosen. Then both $w_i(u,v)$ and $w_j(u,v)$ were increased by 1. If this step happens to introduce a cycle in j's memory web, the weakest arc possible was deleted to make $G_j$ free of cycles (corresponding to j reevaluating its memory to make it consistent). After a sweep through all agents i (and subsequent communication), all familiarities were decreased by φ and all memories by μ (for all our simulations we consider the case φ = μ and drop μ). If this degradation of F or w returns negative values, the value was set to zero. Finally, to mimic misunderstandings, misinterpretations, lies etc., noise was added to the system by rewiring arcs with a expectation value of ρ times per time step. These steps are illustrated in Fig. 2.

## Group formation

The positive feedback from communication, in the real world as well as our model, strengthens ties between similar people, making similar people more similar. In other words, communication plays a role in establishing social groups or communities. Communication also makes the memories of individuals who communicate frequently more similar. In Fig. 3, we show a snapshot configuration in one of our simulation runs. Agents are more strongly connected to each the more similar their memory webs are. Here, similarity is measured by the Jaccard index [18] J—the number of arcs in common to the memories of the two individuals $|A_i \cap A_j|$ divided by the total number of arcs in common to the two individuals $|A_i \cup A_j|$. In Fig. 4 we show a dynamic picture of how groups form and split over the evolution of the simulation.

A more quantitative picture of the group structure is seen in Fig. 5 where we investigate the response of the system to the communication rate. In Fig. 5A we graph the average number of groups detected by a cluster identification algorithm g (see Methods) as a function of the communication rate for different values of the memory and acquaintance degradation rate φ. For larger φ-values, g has a peak for intermediate communication rates. This is, as we will see below, a regime where the population groups into communities that have similar memories within a community but different between different communities. We call this a *diversity regime*. In the Supporting Information, we show that this observation is true also for weak misunderstanding rates. In Fig. 5B we show the average Jaccard indices between the memory webs of pairs of agents both restricted to pairs within the same community and all pairs. At high communication rates the system conforms to society of only one cluster (g = 1 in Fig. 5A) with almost complete consensus among the agents (J is close to one in Fig. 5B). For low communication rates, errors and memory degradation are more influential on the system's behavior than communication—the system will be otherwise
6

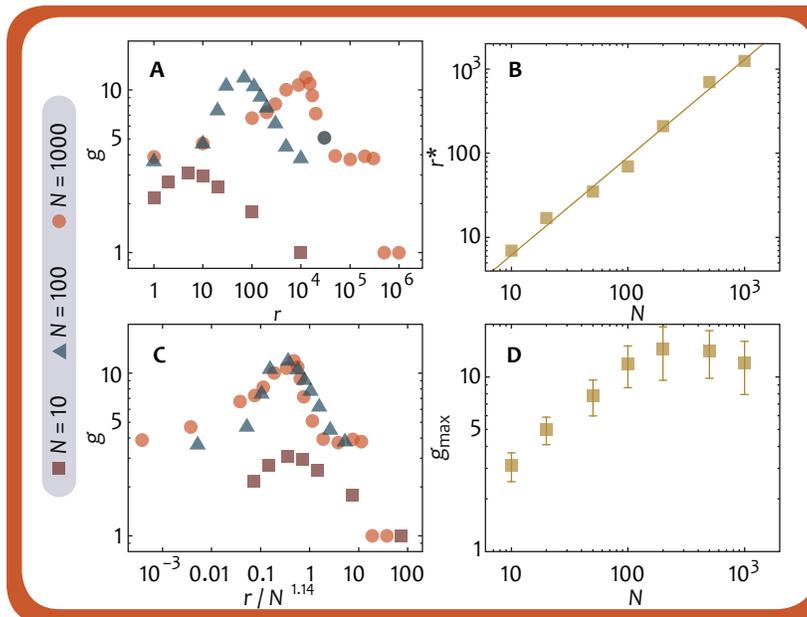

**Fig. 6.** Scaling relationships in the community structure and collective memories. In panel (A), we plot the number of groups $g$ for various system sizes as functions of the communication rate. (B) shows the $r$-values maximizing $g(r)$, $r^*$, as a function of $N$. The solid line follows a power-law with exponent 1.14 (i.e. a functional form $N^{1.14}$), a value obtained from linear regression. (C) displays the number of groups against $r / N^{1.14}$, confirming the size-scaling of the peak's position. (D) shows $g_{max} = g(r^*)$ as a function of the number of agents. Parameter values are $\rho = 0$ and $\varphi = 0.3$. The values are sampled after 440 time steps (when the system has reached equilibrium). We use 50 averages. Errorbars are smaller than the symbol size in all panels except (D).

almost random, which is manifested in low Jaccard indices. The similarity within groups is larger than across the population as a whole, and this difference is largest around the peak of $g$ (note the logarithmic scale, the difference is around ten times larger). A little counter-intuitively, $J$ decreases to a minimum as $r$ increases (Fig. 5B). The reason is not that the number of overlapping association arcs decrease, but that the average number of associations per pair of individuals, $\langle |A_i \cup A_j| \rangle$, grows faster than the number of overlapping arcs $\langle |A_i \cap A_j| \rangle$ (Fig. 5C). Eventually, as $r$ increases, $\langle |A_i \cap A_j| \rangle$ starts growing faster than $\langle |A_i \cup A_j| \rangle$ which makes $J$ grow again. In a real system this corresponds to that, as people communicate more, narratives first spread in the population before they nucleate to communities. The most important conclusion from Fig. 5 is perhaps that both relatively high communication rates and memory degradation are needed for the diversity regime to appear—if people only communicate and do not forget, a uniform collective memory will emerge.

To further investigate the behavior of our model, we look at how the number of groups (the diversity regime) scales with the communication rate for different system sizes (Fig. 6A). First we notice that there is a peak (characteristic of a diversity regime) present for all the three sizes we investigate. The peak's position moves to the right with increasing $N$, meaning, quite naturally, that the larger the system is, the more the agents have to communicate in order for a global consensus to emerge. This is to say, if there is consensus in a subsystem and, as long as the conversation rate is conserved, a diversity regime will be inevitable in a larger context. The size scaling of the peak $r$-value, $r^*$, is directly investigated in Fig. 6B. Where we find (for no rewiring and $\varphi = 0$) that $r^*$ is proportional to $N^{1.14(4)}$.* This means the boundary of the diversity regime increases faster than linear (as a function of $N$)—to keep consensus in a twice as large system one would need to communicate 2.6 times more. For groups of people trying to maintain a collective memory by building museums, monuments, memorial days, ceremonies, etc. [19], this result suggests that their goal will be hard to achieve in a growing population. This observation in Fig. 6C where we plot $g$ against an efficient, rescaled communication rate and find the peaks located at the same $r$-value. In Fig. 6D we look at how the maximum number of groups $g_{max}$ scales with $N$. This curve seems to converge to a value around 10 as $N$ grows. For systems with misunderstanding ($\rho > 0$), $g_{max}(N \to \infty)$ is larger. The fact that $g_{max}$ does not grow without limits would mean that the average size of the groups will be proportional to $N$ in the large-$N$ limit—if the system grows then so does the communities, with the same rate.

Finally, (not shown) we note that cycles, inconsistencies, appear in our simulations in the regime of high communication rates $r \geq 1000$ and high error rates $\rho \geq 1000$ ($\varphi = \mu = 0$).

## Discussion

The concept of memory has both psychological and social dimensions [2]. In historical discourse, it has lately drifted more towards the latter [20]. We model collective memory based on five principles—the experience of actual events, communication across social networks, reinforcement of both memories and social ties from the communication, errors and misconceptions, and forgetting. Our model takes an external memory web as input, a network meant to represent a hypothetical web of associations of unbiased, well informed but otherwise normal individuals. We use an empirical dataset for this seed network constructed from the life-stories of fourteen Chinese villagers. This dataset says something about memory webs in its own right. Most of the events are connected into a large component—the villagers can associate one event to another and thereby cover most of the important events around them during their lives (Ref. [12] discusses how one define historical "cases" relatively self-consistent sequences of associations of events, from this dataset). Most of the events are conceived as affected by and affecting other events. The in- and outdegrees are right-skewed, not far from power-law distributed, meaning that most of the events are relatively minor, but a fraction of them are very much more influential than the average. Still there does not seem to be a strong hierarchical organization of the memory webs so there can always appear arbitrarily influential events. Furthermore, the dataset is clustered into groups of events with relatively weak coupling relations between each other. These clusters are also possible "cases" in the sense of Ref. [12].

---

*The value is obtained by a maximum likelihood method. The number in the parenthesis gives the standard error.

The model predict three different scenarios of collective memories—either the memories are very personal (each agent has its own view of history), or all agents think more or less the same, or (the intermediate case) there are distinct groups which share the same view of history. The latter "diversity regime" with a community structure in terms of memory is also reflected in the social network—it too shows groups which are densely connected between and weakly connected between each other. An example of such a diversity regime in society could be the Israeli and Palestinian diverging textbook accounts of history of the state of Israel [7]. For an increasing population, the diversity regime will require higher communication rates to appear—it thus seems quite unavoidable, at least if one does not take into account that the human population is bounded. At the same time, the number of different collective memories is bounded so that the distinct groups will become larger rather than more as the population grows. If one translates memories to attitudes or norms in general similar observations have been done in e.g. Ref. [21] where the authors discuss why global polarizations of attitudes are rare but do exist.

The potential contributions of this work are twofold: it provides a comprehensive framework for the exploration of the basic mechanisms that underlie collective memory construction and re-construction, and it offers a quantitative set of tools for the characterization of large-scale patterns of collective memories. Moreover, we believe that this work is only the beginning of mechanistic models of the influence of communication on memory. We expect that further exploration on the interplay between autobiographical memory construction and processes of social interaction will provide critical information on how we go about establishing our sense of personal identity and how we come to share a common understanding of history with others. An open question is the role of the events of largest degree, could these function as breaking points not only in the lives of people but also the emergent group structures of collective memory?

## Methods

### Degree distribution

We used a method proposed by Clauset et al. [22] to analyze the probability distributions of degree. This method takes any empirical frequency distribution as its input, and then uses maximum-likelihood estimation to obtain the parameter values—the exponent and the low-degree cut-off (under which the power-law scaling is not required to hold). Then the method generates a number, in our case 1000, of synthetic power-law distributions that can be compared to the original data, using the Kolmogorov–Smirnov (KS) statistic. If one finds, by the KS statistic, that less than 10% of the synthetic data sets are deemed compatible with (in the sense that they could be drawn from the same distribution) the empirical data, then one can exclude the possibility that the empirical data is power-law distributed.

### Clustering coefficient

The clustering coefficient is a measure of the density of triangles in the network. Technically we define the clustering coefficient of a network as [23]

$$C = \frac{3 \times \text{number of triangles}}{\text{number of connected triples}} \quad (1)$$

where a connected triple is a subgraph of three vertices and two arcs, and the factor three normalizes the coefficient to the interval [0,1]. The clustering coefficient of a vertex, measuring the relative number of triangles the vertex participates in is defined as

$$C_i = \frac{T_i}{\binom{k_i}{2}} \quad (2)$$

where $T_i$ is the number of triangles $i$ is a part of (and the binomial factor is the maximal number of $T_i$ given $i$'s degree).

### Degree correlations

Just like the clustering coefficient, we measure degree correlations both as a number and a function of the degree. Plotting the average neighbor degree as a function of the degree gives more detailed information, the assortativity $R$ is a more succinct approach, returning just a number [23]. Technically, the assortativity is Pearson's correlation coefficient of the degree at the ends of an arc.

$$R = \frac{\text{Cov}(k_{\text{from}}, k_{\text{to}})}{\sigma(k_{\text{from}})\,\sigma(k_{\text{to}})} \quad (3)$$

where the numerator is the covariance of the degree at the vertex the arc originates at and the degree of the vertex at the vertex the arc lands at, and the denominator consists of the standard deviations of these two degrees. This measure ranges from –1 to 1 with negative values representing negative degree correlations (so low-degree vertices typically are connected to large-degree vertices), and positive values indicates a positive degree–degree correlation so that vertices of a large degree are connected to each other and vertices of low degrees are connected to other low-degree vertices.

### Reference model

To put the clustering coefficient and assortativity into context we compare the values to average over a random model where the sizes (number of vertices and arcs) are fixed as are the set of degrees (so that there are e.g. as many vertices with indegree 1 and outdegree 3 in the model graphs as the original graph). This model can be sampled by going through the arcs iteratively and for any arc choose another arc randomly and swap the vertices pointed to by the arcs. This method, proposed by Gale [24] and popularized by Sneppen & Maslov [25] and Shen-Orr et al. [26], assumes the degree distribution to be the most fundamental network structure and investigates what other structural biases there are in the network, except the degree.

### Community structure

To investigate the both the community (or cluster) structure of the empirical network and the memory web we use the method of Ref. [27]. This algorithm sets out by a defining a measure of the modularity $Q$ of a partition of a weighted graph into subgraphs. For comparing the similarity of memory webs $Q$ is defined as:

$$Q = \frac{1}{2J} \sum_{ij} \left( \Lambda_{ij} - \frac{\Lambda_i \Lambda_j}{2\Lambda} \delta_{ij} \right), \quad (4)$$

where $\Lambda_{ij}$ is the fraction of the total Jaccard similarity $J$ between two partitions $i$ and $j$, $\Lambda_i$ is the $i$'th row (or column) sum and $\Lambda$ is the total weights (or, equally, the sum of over all partitions). $\Lambda_{ij}$ is 1 if $i$ and $j$ are in the same partition, 0 otherwise. The first term of the sum, $\Lambda_{ii}$, the total weight within the same partition should be as large as possible in a partitioning reflecting the community structure of the system. The second term is the expectation value of, $\Lambda_{ii}$, in a random network given the set of degrees in the partitions of the current partitioning. A positive $Q$ means that there are more arcs between vertices of the same partition and fewer arcs between vertices of different partitions than expected in a random partitioning. The method then maximizes $Q$ over all partitionings (which is intractable to do exactly but can be done approximately with a fairly high accuracy). In this paper we are primarily interested in $g$, the number of partitions in the partitioning maximizing $Q$.

For the empirical memory web we calculate $q$ by treating the network as undirected and unweighted. Then $q$ can be written

$$q = \max \sum_i \left( e_{ii} - \left( \sum_j e_{ij} \right)^2 \right) \quad (5)$$

where $e_{ij}$ is the fraction of arcs going from partition $i$ to partition $j$, and the maximization is over all partitions.

### Acknowledgments

The authors thank James Moody for the data and Martin Rosvall for help with visualizing the time evolution of clusters. This research was supported by the Wenner–Gren Foundations (SL), the Swedish Foundation for Strategic Research (PH), the Swedish Research Council (PH), and the WCU program through NRF Korea funded by MEST R31–2008–000–10029–0 (PH).

## Supplementary Information

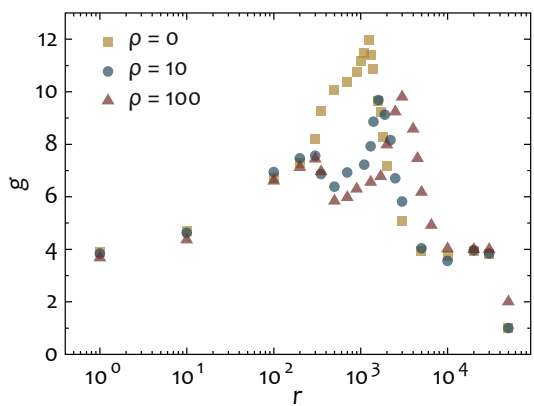

In this Supplementary Information, we present a picture showing the stability of Fig. 4A to the misunderstanding parameter $\rho$. We use $\varphi = 0.3$ and keep the other parameter values the same as in Fig. 4. The error bar is smaller than symbol size. If there is misunderstanding, or noise, in the system the maximum number of groups is lower and the peak is split in two, the larger peak happens at larger $r$-values. The main quantitative conclusion from the study at $\rho = 0$, the existence of a diversity regime, remains. The split peak, we believe, is an effect of the communities nucleating more heterogeneously. Around $r = 500$ there are some stronger communities that takes more of the links than in the noiseless case, whereas other groups, that would be communities without noise, the gets grouped into one. The effect of this would, we believe, be the observed dip in the $g(r)$-curve around $r = 500$. Yet increasing communication makes the system able to overcome the noise so that the peak is almost recovered although displaced toward larger $r$.